\documentclass[draft,showpacs,eqsecnum,nofootinbib,aps]{revtex4}
\renewcommand{\theequation}{\arabic{equation}}
\def\beq{\begin{equation}}
\def\eeq{\end{equation}}
\def\bea{\begin{eqnarray}}
\def\eea{\end{eqnarray}}

\begin{document}
\title{COSMIC STRING SPACETIME IN DILATON GRAVITY \\
      AND FLAT ROTATION CURVES}
\author{Tae Hoon Lee}
\email{published in Modern Physics Letters A19 (2004) 2929}
\affiliation{Department of Physics, Soongsil University, Seoul
156-743 Korea}
\date{\today}%
\begin{abstract}
In dilaton gravity theories, we consider a string-like topological
defect formed during $U(1)$ gauge symmetry-breaking phase
transition in the early Universe, and far from the cosmic string
we have vacuum solutions of the generalized Einstein equation. We
discuss how they can be related to the flatness of galactic
rotation curves.
\end{abstract}
\maketitle

\section{Introduction}
\setcounter{equation}{0}
\renewcommand{\theequation}{\arabic{section}.\arabic{equation}}
   Topological defects of various kinds could be generated
when symmetry-breaking phase transitions took place in the early
Universe. Defects such as global monopoles or cosmic strings were
thought of as the seeds for galaxy and large-scale structure
formation,\cite{Bennet} and their remnants might remain as
galactic dark matter.

The observed flatness\cite{flat} of spiral galaxy rotation curves
suggests the existence of dark matter whose energy density goes
like $1/r^2$. Since the global monopole which was found by
Barriola and Vilenkin\cite{vi} as a solution of Einstein's
equations has energy density proportional to $1/r^2$, it was
suggested by Nucamendi and others\cite{nuc} that the global
monopole could account for some fraction of the galactic dark
matter. Even though Harari and Loust\'{o}\cite{ha} noticed that
the monopole core mass is negative and that there are no bound
orbits, Nucamendi {\it et} {\it al.}\cite{nuc} showed that there
is an attractive region where bound orbits exist, by the
introduction of a nonminimal coupling of gravity to the global
monopole. Various, generalized scalar fields coupled to gravity
with minimal coupling\cite{scalar} and with nonminimal
coupling\cite{BT} were also studied in Einstein's theory as
candidate dark matter. On the other hand, Banerjee {\it et} {\it
al.}\cite{nuc} noted that the global monopole in Brans-Dicke
theory also exerts gravitational pull on a test particle moving in
its spacetime. Recently we thus have studied in more detail the
global monopole in Brans-Dicke theoy and discussed the importance
of the ${\ln} \,r$ term of a gravitational potential in its
possible relation to flat rotation curves.\cite{TB} Such $\ln \,r$
terms can appear in various other cases\cite{THBM} including
global cosmic strings.\cite{V}

The cosmic string is another solution of Einstein's equations in
the effectively (2+1)-dimensional spacetime. It was found about 20
years ago and has been extensively studied\cite{V} because of its
possibly important role in the large-scale structure
formation.\cite{Bennet} Since the energy density generated by long
global strings varies\cite{V} as $1/r^2$ as global monopoles and
it thus was thought to be responsible for the flatness of rotation
curves in quasars,\cite{Miramare} we suggest it would be related
to the flatness of rotation curves in spiral galaxies also. Even
if a gauge cosmic string induces only a deficit angle\cite{V} in
Einsten's theory of gravity, it makes surrounding spacetime curved
in scalar-tensor theories. Cosmic strings in various scalar-tensor
gravity theories including dilaton gravity have been studied by
many authors.\cite{small}

In dilaton gravity theories, we consider a $U(1)$ gauge cosmic
string formed in the early Universe and have cylindrically
symmetric solutions of field equations far from the cosmic string.
By comparing this vacuum solutions with the perturbative solutions
obtained by Gregory and Santos,\cite{small} we determine the
spacetime metric around the string. Applying the formula (3.23)
for rotation velocities to the solutions, we get a constant value
of rotation velocities of stars in a galaxy seeded by the string,
$v_{rot.}^{(s)} \propto G\eta^2 \sim 10^{-6} $ which is to be
given in Eq. (3.28), when we consider a cosmic string formed at
the GUT scale of about $ 10^{16}$ GeV. This constant value
obtained with a single string looks too small to explain the
measured value\cite{flat} of rotation velocities in galactic
halos. However, the value of rotation velocities around a bundle
of a few hundred cosmic strings might be comparable to the
measured value.\cite{flat} Even though such multi-string structure
surrounded by the outer void region could be constructed by the
possible force\cite{CQ} of attraction between parallel strings or
by the process of string accumulation,\cite{accu} we present a
more concrete field-theoretical two-dilaton model, in illustration
of the domain-wall structure\cite{domain} of strings. The flatness
of spiral galaxy rotation curves might be due to the constant
value of rotation velocities around the remnants of the
multi-string, which has been possibly preserved because of the
scaling property of the cosmic string network.\cite{scaling} On
the other hand, it can be also realized if we consider a cosmic
string formed near the string-unification scale.\cite{ms}

The rest of the paper is organized as follows. In Sec. 2, we
derive equations of motion for $U(1)$ gauge fields $Z_\mu$ and a
scalar field $\Phi^0$ coupled to dilaton gravity, and with
appropriate ansatzes we reduce the equations to the cylindrically
symmetric forms when the $U(1)$ symmetry is spontaneously broken.
In Sec. 3, we find vacuum solutions far from the $Z$-string formed
at the origin. We calculate the rotation velocities of objects
circulating the cosmic string and discuss their relation to flat
rotation curves in galaxies. In Sec. 4, we present a two-dilaton
model for the flatness of rotation curves in the galactic halo
region. The summary and discussion is given in Sec. 5.

\section{Z-strings in dilaton gravity theories}
\setcounter{equation}{0}
\renewcommand{\theequation}{\arabic{section}.\arabic{equation}}

Since Jordan, Brans and Dicke\cite{BD} introduced a scalar field
instead of $G^{-1}$ for the purpose of generalizing Einstein's
theory of gravity to incorporate Mach's principle, various
scalar-tensor theories of gravity have been studied. A
four-dimensional dilaton gravity obtained as a low energy
effective theory from fundamental strings also has the form of a
scalar-tensor theory, with non-trivial couplings of the dilaton to
matter and a possible dilaton potential. When a potential of the
dilaton field, $e^{-2\phi} V (\phi)$, is included, an action of
dilaton gravity theories is given by
\begin{equation}
S =\frac{1}{16\pi G}\int d^4 x
\sqrt{-\tilde{g}}e^{-2\phi}[\tilde{R} + 4 {\tilde{g}}^{\mu
\nu}\partial_{\mu}\phi \partial_{\nu} \phi - e^{-2\phi} V(\phi)]
+S_m
\end{equation}
with the action for dilaton-matter couplings, $S_m =\int d^4 x
\sqrt{-\tilde{g}}e^{2a\phi}{\cal L}_m $.

Let us consider a theory of gauge fields and a scalar field with a
broken $U(1)$ gauge symmetry in the dilaton gravity. By the
conformal transformation
\begin{equation}
\tilde{g}_{\mu\nu}=e^{2\phi}g_{\mu\nu},
\end{equation}
from the above action we have the action of the theory in the
socalled Einstein conformal frame.\cite{da}
\begin{equation}
S =\frac{1}{16\pi G}\int d^4 x \sqrt{-{g}}[ R
 - 2 g^{\mu \nu}\partial_{\mu}\phi \partial_{\nu} \phi
 -V (\phi) ] +S_m [ e^{2\phi} g_{\alpha \beta}, Z_{\mu}, \Phi^0],
\end{equation}
where
\begin{eqnarray}
S_m [e^{2\phi} g_{\alpha \beta}, Z_{\mu}, \Phi^0] =&&\int d^4 x
\sqrt{-g}e^{2(a+2)\phi}{\cal L}_m (e^{2\phi} g_{\alpha \beta}, Z_{\mu}, \Phi^0 ) \nonumber\\
= &&-\int d^4 x
\sqrt{-g}[\frac{1}{4}e^{2a\phi}Z^{\mu\nu}Z_{\mu\nu}
+e^{2(a+1)\phi}g^{\mu\nu}(D_{\mu}\Phi^0)^{\dagger}D_{\nu}\Phi^0 \nonumber\\
&&+e^{2(a+2)\phi} \, V_\eta (\Phi^0)].
\end{eqnarray}
Here $Z_\mu$ is a neutral gauge field and $Z_{\mu\nu}=\partial_\mu
Z_\nu -\partial_\nu Z_\mu$ are components of its field strength
with $Z^{\mu\nu}=g^{\mu\alpha}g^{\nu\beta}Z_{\alpha\beta}$.
$\Phi^0$ is the scalar field, and its covariant derivative and
potential are given by
\begin{eqnarray}
D_\mu \Phi^0 =&&(\partial_\mu +ie Z_\mu)\Phi^0,\nonumber\\ V_\eta
(\Phi^0)=&&\lambda(\vert \Phi^0 \vert^2 -\eta^2)^2.
\end{eqnarray}

Varying the action (2.3) with respect to the fields, we have
following equations: The equation for the gauge fields $Z_\mu$ is
\begin{equation}
\frac{1}{\sqrt{-g}}\partial_{\nu} e^{2a\phi} \sqrt{-g}
g^{\mu\alpha} g^{\nu\beta} Z_{\alpha\beta}=i e \,
e^{2(a+1)\phi}g^{\mu\nu}
[{\Phi^0}^{\dagger}(D_{\nu}\Phi^0)-(D_{\nu}\Phi^0)^{\dagger}{\Phi^0}].
\end{equation}
The equation for the scalar field $\Phi^0$ is given by
\begin{equation}
\frac{1}{\sqrt{-g}}D_{\nu}^{\dagger} e^{2(a+1)\phi} \sqrt{-g}
g^{\mu\nu}(D_{\nu}\Phi^0)^{\dagger}=2\lambda e^{2(a+2)\phi}(\vert
\Phi^0 \vert^2 -\eta^2){\Phi^0}^{\dagger},
\end{equation}
and the equation for the dilaton field $\phi$
\begin{equation}
\frac{1}{\sqrt{-g}}\partial_{\mu} \sqrt{-g}
g^{\mu\nu}\partial_{\nu}  \phi
 -\frac{1}{4}\frac{\partial V}
{\partial \phi}=  -\frac{4\pi G}{\sqrt{-g}}\frac{\delta
S_m}{\delta \phi} ,
\end{equation}
where
\begin{eqnarray}
\frac{1}{\sqrt{-g}} \frac{\delta S_m}{\delta \phi }
&&=-[2(a+2)e^{2(a+2)\phi}\lambda(\vert \Phi^0 \vert^2 -\eta^2)^2
\nonumber \\
&&+2(a+1)e^{2(a+1)\phi}g^{\alpha\beta}(D_{\alpha}\Phi^0)^{\dagger}
D_{\beta}\Phi^0    + \frac{a}{2} e^{2a\phi}
Z^{\alpha\beta}Z_{\alpha\beta}].
\end{eqnarray}
Einstein's field equations are given by
\begin{equation}
G_{\mu \nu} =8\pi G T_{\mu \nu}+2\partial_\mu \phi
\partial_\nu \phi
-\frac{1}{2}g_{\mu\nu} \{ 2g^{\alpha\beta}\partial_\alpha \phi
\partial_\beta \phi+ V (\phi) \},
\end{equation}
where the energy-momentum tensor
\begin{eqnarray}
T_{\mu \nu} &&\equiv -\frac{2}{\sqrt{-g}} \frac{\delta S_m}{\delta
g^{\mu \nu} }  \nonumber\\ &&= -  [g_{\mu\nu} \{
e^{2(a+2)\phi}\lambda(\vert \Phi^0 \vert^2
-\eta^2)^2+e^{2(a+1)\phi}g^{\alpha\beta}(D_{\alpha}\Phi^0)^{\dagger}
D_{\beta}\Phi^0  \} \nonumber \\
&&-2e^{2(a+1)\phi}(D_{\mu}\Phi^0)^{\dagger}D_{\nu}\Phi^0
+e^{2a\phi}(\frac{1}{4}g_{\mu\nu} Z^{\alpha\beta}Z_{\alpha\beta}
-g^{\alpha\beta}Z_{\alpha\mu}Z_{\beta\nu})].
\end{eqnarray}

With the potential of the field $\Phi^0$ given by $ V_\eta
(\Phi^0)=\lambda(\vert \Phi^0 \vert^2 -\eta^2)^2$ in Eq. (2.5), we
are interested in a string-like defect situated at the origin, and
we thus take ansatzes for the scalar field and the gauge field
as\cite{Linet}
\begin{eqnarray}
\Phi^0 =\eta f(\rho) e^{i n\varphi}, \nonumber \\ Z_\mu
=\frac{1}{e}[P(\rho)-n]\delta_\mu^\varphi ,
\end{eqnarray}
and for the spacetime metric with cylindrical symmetry
as\cite{boost}
\begin{equation}
ds^2
 = g_{\mu\nu}dx^\mu dx^\nu =A(\rho)(-dt^2 + dz^2)
+ d\rho^2 + B(\rho) d\varphi^2.
\end{equation}

Defining new, dimensionless variable and quantities as
\begin{eqnarray}
\sqrt{\lambda}\eta\rho=r, \nonumber\\
\frac{e^2}{\lambda}=\alpha, \nonumber \\ 8\pi G \eta^2 =\kappa,\nonumber \\
\lambda\eta^2 B \rightarrow B ,
\end{eqnarray}
we have following equations from Eqs. (2.6)-(2.10), when
$V(\phi)=\lambda\eta^2 {\cal V}(\phi)$. The gauge field equation
(2.6) reads
\begin{equation}
\frac{1}{\Psi^a A B^{-\frac{ 1}{2}}} ( \Psi^a AB^{-\frac{ 1}{2}}
P')'=2\alpha \Psi f^2 P.
\end{equation}
The scalar field equation (2.7) becomes
\begin{equation}
\frac{1}{\Psi^{a+1}A B^{\frac{1}{2}}}( \Psi^{a+1}AB^{\frac{1}{2}}
f')'=2\Psi (f^2-1)f+\frac{fP^2}{B}.
\end{equation}
The dilaton equation (2.8) becomes
\begin{equation} \frac{1}{A
B^{\frac{1}{2}}}( AB^{\frac{1}{2}} \phi')'
-\frac{1}{4}\frac{\partial{\cal V}}{\partial \phi}
=\kappa\Psi^a[\frac{a}{2}\frac{P'^2}{\alpha
B}+(a+2)\Psi^2(f^2-1)^2 +(a+1)\Psi(f'^2 +\frac{f^2P^2}{B})].
\end{equation}
The ($\rho \rho$)-component of the Einstein equation (2.10) reads
\begin{equation}
 \frac{1}{2}(\frac{A'}{A})^2
+\frac{A'}{A}\frac{B'}{B}+{\cal V}-2(\phi')^2 =\kappa\Psi^a [2\Psi
f'^2 +\frac{P'^2}{\alpha
B}-2\Psi^2(f^2-1)^2-2\Psi\frac{f^2P^2}{B}].
\end{equation}
The other components of the Einstein equation are reduced to the
following two equations.
\begin{eqnarray}
 \frac{A''}{A} +\frac{1}{2}\frac{A'}{A}\frac{B'}{B}+{\cal V}
&&=\kappa\Psi^a[\frac{P'^2}{\alpha B}-2\Psi^2(f^2-1)^2 ], \\
 \frac{B''}{B}
 +\frac{A'}{A}\frac{B'}{B}-\frac{1}{2}(\frac{B'}{B})^2+{\cal V}
&&= -\kappa\Psi^a[\frac{P'^2}{\alpha B}+2\Psi^2(f^2-1)^2+4
\Psi\frac{f^2P^2}{B}],
\end{eqnarray}
where $\Psi\equiv e^{2\phi}$ and $\frac{\partial}{\partial r}A =A'
, ...$ have been used for the simplicity. Above equations
(2.15)-(2.20) are consistent with the result of Verbin {\it et
al.}\cite{class} when ${\cal V}\equiv 0$.

\section{Vacuum solutions and circular motion around cosmic string}
\setcounter{equation}{0}
\renewcommand{\theequation}{\arabic{section}.\arabic{equation}}

Let us consider the case ${\cal V}= 0$ in this section. Far from
the $Z$-string, $r>>1$, the vacuum solutions
($\vert\Phi^0\vert\simeq\eta$) are given with the following
boundary conditions.\cite{Linet}
\begin{equation}
f\simeq 1,\,\,\, P\simeq 0,
\end{equation}
which give us, from Eqs. (2.15) and (2.16),
\begin{equation}
\Psi^a AB^{\frac{- 1}{2}} P'=c_p,\,\,\,\Psi^{a+1}AB^{\frac{1}{2}}
f'=c_f
\end{equation}
with constants $c_p$ and $c_f$. In case\cite{Linet} of
\begin{equation}
c_p =c_f =0,
\end{equation}
we have
\begin{equation}
f'\simeq 0,\,\,\, P'\simeq 0,
\end{equation}
which are consistent with Eq. (3.1). With these, Eq. (2.17) gives
us
\begin{equation}
AB^{\frac{1}{2}} \phi'=c_\phi
\end{equation}
with a constant $c_\phi$.

The Einstein equations (2.18)-(2.20) read
\begin{eqnarray}
\frac{1}{2}(\frac{A'}{A})^2
&&+\frac{A'}{A}\frac{B'}{B}-2(\phi')^2 =0,\\
\frac{A''}{A} &&+\frac{1}{2}\frac{A'}{A}\frac{B'}{B}
=0, \\
\frac{B''}{B} &&
+\frac{A'}{A}\frac{B'}{B}-\frac{1}{2}(\frac{B'}{B})^2 = 0.
\end{eqnarray}
We can satisfy Eq. (3.7) with
\begin{equation}
A'=c_a B^{-\frac{1}{2}},
\end{equation}
and we have the equation
\begin{equation}
B'=c_b B^{\frac{1}{2}} A^{-1}
\end{equation}
with the relation between constants, $ c_\phi^2
=\frac{1}{4}c_a(c_a +2c_b)$. With Eqs. (3.9)-(3.10), Eq. (3.8)
reads
\begin{equation}
\frac{B''}{B} +(\frac{c_a}{c_b} -\frac{1}{2})(\frac{B'}{B})^2 = 0,
\end{equation}
whose solution is
\begin{equation}
B=B_i \, r^{2-\epsilon}
\end{equation}
where
\begin{equation}
\epsilon \equiv \frac{4 \frac{c_a}{c_b}}{1+2\frac{c_a}{c_b}}.
\end{equation}
The solutions for another metric coefficient $A$ and the dilaton
field $\phi$ are given by
\begin{eqnarray}
 A(r)&&=A_{i} \,r^{\frac{\epsilon}{2}}, \nonumber\\
 \phi(r)&&=\phi_{i}   + c_{i} (\epsilon)\, {\rm ln}\frac{r}{r_c},
\end{eqnarray}
with $A_{i} =B_{i}^{-\frac{1}{2}}(c_a +\frac{c_b}{2})$ and $c_{i}
(\epsilon) =\sqrt{\frac{1}{2}\epsilon(1-\frac{3}{8}\epsilon)}$.
With these solutions the spacetime metric becomes
\begin{equation}
ds^2 =A_i r^{\frac{\epsilon}{2}}(-dt^2 + dz^2) + dr^2 + B_i
r^{2-\epsilon} d\varphi^2.
\end{equation}

To derive the formula for rotation velocities of objects
circulating the string formed at $r=0$, we consider the following
spacetime metric
\begin{equation}
ds^2
 = g_{\mu\nu}dx^\mu dx^\nu =A(r)(-dt^2 + dz^2)
+ dr^2 + B(r) d\varphi^2,
\end{equation}
which is obtained from the spacetime (2.13) with appropriate
reparametrizations.\cite{small} Using the definition
\begin{equation}
(  \frac{dt}{d\tau}, \, \frac{dr}{d\tau},\, \frac{dz}{d\tau} ,\,
\frac{d\varphi}{d\tau} ) \equiv (\dot{t},\, \dot{r},\, \dot{z},\,
\dot{\varphi})
\end{equation}
with the proper time $\tau$, from Eq. (3.16) with $\dot{z}=0$ we
get the equation for the Lagrangian:
\begin{equation}
{\cal L}(r, \dot{r}, \dot{\varphi}, \dot{t}; \tau)\equiv -A(r)
{\dot{t}}^2 + {\dot{r}}^2 + B(r){\dot{\varphi}}^2 =-1
\end{equation}
where it is used that $ds^2 =-d\tau^2 $ in our unit system. Since
${\partial {\cal L}}/{\partial \varphi}=0$ and ${\partial {\cal
L}}/{\partial t}=0$, we have the constants of motion
\begin{equation}
\frac{\partial {\cal L}}{\partial \dot{\varphi}} \equiv  2 B
\dot{\varphi}=2L,  \,\,\, \frac{\partial {\cal L}}{\partial
\dot{t}} \equiv  -2 A \dot{t}=-2E.
\end{equation}

When $\dot{z}=0$, the geodesic equation in the spacetime (3.16),
with the help of the above equations, reads
\begin{equation}
{\dot{r}}^2 +V_{eff.}(r)=0,
\end{equation}
where
\begin{equation}
V_{eff.}(r)=1+\frac{L^2}{B(r)} -\frac{E^2}{A(r)}.
\end{equation}
We require the following conditions for the objects to have
circular motions:\cite{phi}
\begin{equation}
\dot{r}=0, \,\,\, \frac{\partial V_{eff.}}{\partial r}=0, \,\,\,
\frac{\partial^2 V_{eff.}}{\partial r^2} > 0.
\end{equation}
Following the same procedure as Refs. [23]-[24], we solve the
above equations, express $\dot{\varphi}$ and $\dot{t}$ as
functions of metric coefficients, and have the formula for the
rotation velocity
\begin{equation}
v_{rot.} \equiv \frac{B^{\frac{1}{2}}}{A^{\frac{1}{2}}} \frac{ d
\varphi}{ dt}=\sqrt{\frac{A'B}{AB'}}.
\end{equation}

Applying the above equation to Eq. (3.15), we obtain the circular
velocity of the objects around the string,
\begin{equation}
v_{rot.}
=\sqrt{\frac{c_a}{c_b}}=\sqrt{\frac{\epsilon}{4(1-\frac{\epsilon}{2})}}.\label{rot}
\end{equation}
As we can see in Eq. (\ref{rot}), the velocities of rotating
objects far away from ($r>>1$) the cosmic string are non-zero and
$r$-independent for $0 \leq \epsilon<2$. To discuss flat rotation
curves in a galaxy whose formation was seeded by the cosmic string
in the early Universe, we think of a situation where $0 <
\epsilon<2$ for $r_c \leq r \leq r_h$ and $\epsilon=0$ for $r_h <
r$, with the radius of the galactic halo $r_h (\simeq 200 - 400
\,{\rm kpc}).$\cite{23} We do not discuss the core region ($ <
r_c$), since it might have very complex structure such as
supermassive black hole.\cite{bh} For the outer void region
($>r_h$) we take the value $0$, instead of $\epsilon$, in Eqs.
(3.12)-(3.15) and have a Minkowskian spacetime as follows.
\begin{eqnarray}
B(r)&&=B_{o} \, r^2, \nonumber \\
 A(r)&&=A_{o},  \nonumber\\
 \phi(r)&&=\phi_{o} ,
\end{eqnarray}
where $B_o =B_i \, r_h^{-\epsilon},\,A_o =A_i \,
r_h^{\epsilon/2}$, and $\phi_o=\phi_i + c_i {\rm ln}(r_h/r_c)$ are
constants of integrations which are determined by the boundary
conditions at $r=r_h$, and $B_o =B_i \, r_h^{-\epsilon}$ can be
related with deficit angles.\cite{V}

Compared by the metric which was, to order $\eta^4$ in units of
the Planck mass, determined by Gregory and Santos\cite{small} and
others,
\begin{equation}
ds^2 ={\hat{r}}^{(a+1)^2 (\hat{\mu}\eta^2/4)^2}(-dt^2
+d{\hat{r}}^2 +dz^2)+(1-\mu/4\pi){\hat{r}}^{2-(a+1)^2
(\hat{\mu}\eta^2/4)^2}d\varphi^2,
\end{equation}
the parameter $\epsilon$ can be given as
\begin{equation}
\epsilon  =\frac{(a+1)^2 \frac{{\hat{\mu}}^2\eta^4}{8}}{1+ (a+1)^2
\frac{{\hat{\mu}}^2\eta^4}{32}  }.
\end{equation}
With the string energy per unit length $\mu$ and ${\cal O}(1)$
quantity $\hat{\mu}\equiv \mu/2\pi\eta^2$, we have
coordinate-transformed here as ${\hat{r}}^{1+{(a+1)^2
{\hat{\mu}}^2 \eta^4}/{32}}=r $ in comparison with Eq. (3.15), and
from Eq. (3.24) we get the velocity formula of objects rotating
the single cosmic string
\begin{equation}
v_{rot.}^{(s)}\simeq \frac{\vert a+1
\vert\hat{\mu}}{4\sqrt{2}}\eta^2 ,
\end{equation}
which is useful in the region where the galactic halo exists, $r_c
\le r \le r_h $.

If we consider a cosmic string formed at the GUT scale of
$M_{GUT}\sim 10^{16}$ GeV, then we have a constant value of
rotation velocities of stars in the string-seeded galaxy,
$v_{rot.}^{(s)} \sim 10^{-6} $, from Eq. (3.28) with $a \neq -1$.
This constant value by a single string looks too small to explain
the measured value\cite{flat} ($\gtrsim 3\times 10^{-4}$) of
rotation velocities in galactic halos. However, if a bundle of $N$
cosmic strings formed at $M_{GUT}$ seeded one galaxy, then the
constant value of rotation velocities become much larger as
$v_{rot.} \simeq N v_{rot.}^{(s)}$ and could be comparable to the
measured value in the galactic halos, with $N \gtrsim 300$. The
constant value of rotation velocities can remain from the era of
the pregalaxies, with the help of the scaling solution for the
cosmic string network,\cite{scaling} and it can be related to the
flatness of rotation curves in spiral galaxies. Such a domain-wall
structure of strings explained below Eq. (3.24) can be realized
also in a two-dilaton model, which is to be presented in the
following section.

\section{Two-Dilaton Model}
\setcounter{equation}{0}
\renewcommand{\theequation}{\arabic{section}.\arabic{equation}}

In the last section we have suggested a very simple model for the
explanation of flat rotation curves in the galactic halo, and now
we present a more concrete, field-theoretical model in
illustration of the domain-wall structure\cite{domain} of strings.
Let us introduce another dilaton field $\chi$ in addition to the
$\phi$ field with a $\phi$-$\chi$ interation potential
\begin{equation}
V_{\chi}(\phi,
\chi)=\frac{\lambda_{\chi}}{2}[(\phi-\chi)(\phi-\phi_0 )]^2.
\end{equation}
The action in this model is given by
\begin{equation}
S =\frac{1}{16\pi G}\int d^4 x \sqrt{-{g}}[ R
 - 2 g^{\mu \nu}\partial_{\mu}\phi \partial_{\nu} \phi
 -\frac{1}{2}e^{2\delta \phi}g^{\mu\nu}\partial_\mu \chi \partial_\nu \chi -V_{\chi}(\phi, \chi)
 ]+S_m
\end{equation}
where $S_m$ is as in Eq. (2.4).

Since $\frac{\partial V_\chi}{\partial \chi}=0,\,\frac{\partial
V_\chi}{\partial \phi}=0$ and $V_{\chi}=0 $ for $\phi=\chi$ or
$\phi=\phi_0$, we have two different vacua $\phi=\chi$ or
$\phi=\phi_0$, where we have following equations of motion for
each dilaton field.
\begin{equation}
\frac{1}{\sqrt{-g}}\partial_{\mu} \sqrt{-g}
g^{\mu\nu}\partial_{\nu}  \chi
 +\lambda_\chi (\phi-\chi)(\phi-\phi_0 )^2 =0,
\end{equation}
\begin{equation}
\frac{1}{\sqrt{-g}}\partial_{\mu} \sqrt{-g}
g^{\mu\nu}\partial_{\nu}  \phi
 -\frac{1}{4}\frac{\partial V}
{\partial \phi}=  -\frac{4\pi G}{\sqrt{-g}}\frac{\delta
S_m}{\delta
\phi}+\frac{\delta}{4}e^{2\delta\phi}g^{\mu\nu}\partial_\mu \chi
\partial_\nu \chi.
\end{equation}
The Einstein equation in this two-dilaton model is given by
\begin{eqnarray}
G_{\mu \nu} &&=8\pi G T_{\mu \nu}+2\partial_\mu \phi
\partial_\nu \phi
-\frac{1}{2}g_{\mu\nu}\{ 2g^{\alpha\beta}\partial_\alpha \phi
\partial_\beta \phi+ V_{\chi} (\phi, \chi)\} \nonumber
\\&& +\frac{1}{2}e^{2\delta\phi}\{ \partial_\mu \chi
\partial_\nu \chi
-\frac{1}{2}g_{\mu\nu}(g^{\alpha\beta}\partial_\alpha \chi
\partial_\beta \chi)\},
\end{eqnarray}
which is same as Eq. (2.10), except its ($\rho\rho$)-component
modified to
\begin{eqnarray}
 \frac{1}{2}(\frac{A'}{A})^2
+\frac{A'}{A}\frac{B'}{B}+{\cal V}-2(\phi')^2
-\frac{1}{2}e^{2\delta\phi}(\chi')^2 &&=\kappa\Psi^a [2\Psi f'^2
+\frac{P'^2}{\alpha B} \nonumber\\&&
-2\Psi^2(f^2-1)^2-2\Psi\frac{f^2P^2}{B}].
\end{eqnarray}

With the same conditions as Eqs. (3.1)-(3.4) in the vacua in the
spacetime (3.16), the $\chi$ field equation (4.3) reads
\begin{equation}
\Psi^\delta AB^{\frac{1}{2}} \chi'=c_\chi
\end{equation}
with a constant $c_\chi$, and the $\phi$ field equation is given
by
\begin{equation}
(AB^{\frac{1}{2}} \phi')'=\frac{\delta}{4} \frac{
c_\chi^2}{\Psi^\delta AB^{\frac{1}{2}}}  .
\end{equation}
Therefore, Eq. (4.6) is reduced to
\begin{equation}
\frac{1}{2}(\frac{A'}{A})^2 +\frac{A'}{A}\frac{B'}{B}-2(\phi')^2
=\frac{1}{2}e^{2\delta\phi}(\chi')^2 ,
\end{equation}
which is given instead of Eq. (3.6), while Eqs. (3.7) and (3.8)
remain same as in the previous section. Eq. (3.7) can be satisfied
with
\begin{equation}
A'=c_a B^{-\frac{1}{2}}
\end{equation}
as in the previous section, and we have the same relation as
before,
\begin{equation}
B'=c_b B^{\frac{1}{2}} A^{-1}.
\end{equation}

In the simple case $\delta=0$, assuming
\begin{equation}
B(r)=B_{in} \, r^2 r^{-\epsilon}
\end{equation}
for $r_c \le r \le r_h $ as Eq. (3.12), we have
\begin{eqnarray}
 A(r)&&=A_{in} \, r^{\frac{\epsilon}{2}}, \nonumber\\
 \phi(r)-\phi_{in} &&=\chi(r)-\chi_{in} =
  \frac{2}{\sqrt{5}} c_{i} (\epsilon)\, {\rm ln}\frac{r}{r_c},
\end{eqnarray}
with $A_{in} =2 c_a B_{in}^{-\frac{1}{2}} /\epsilon$ and
$c_{i}(\epsilon)
=\sqrt{\frac{1}{2}\epsilon(1-\frac{3}{8}\epsilon)}$. If we put
$A_{in}=A_i$, $B_{in}=B_i$, and $\phi_{in} + \frac{2}{\sqrt{5}}
c_i {\rm ln}(r_h/r_c)=\phi_o$, then we have the same domain-wall
structure of strings as in the previous section. However,
$\phi=\chi$ in the galactic halo region ($<r_h$) differently from
the previous section, while $\phi=\phi_o$ in the outer void region
($>r_h$).

\section{Summary and Discussion}

In a spontaneously broken $U(1)$ gauge theory of gauge fields
($Z_\mu$) and a scalar field, coupled to dilaton gravity, we have
derived equations of motion for each fields with appropriate
ansatzes in a cylindrically symmetric spacetime. With the
solutions obtained far away from the $Z$-string formed at the
origin, we have calculated the rotation velocities of the objects
circulating the cosmic string and suggested their relation to flat
rotation curves in the string-seeded galaxy. If the cosmic string
formed at the typical GUT energy scale of $10^{16}$ GeV grew into
the galaxy, then we have a constant value of rotation velocity
around it, $v_{rot.}^{(s)} \propto \eta^2$ ($\sim 10^{-6} $) in
Eq. (3.28), which seems too small. However, the possible
force\cite{CQ} of attraction between parallel strings (or the
process of string accumulation\cite{accu}) could make a bundle of
multi-strings from the cosmic strings, and the constant value of
rotation velocities around the bundle become much
larger\cite{temp} and could be comparable to the measured
one\cite{flat}. The flatness of spiral galaxy rotation curves
might be their remaining property, which has been preserved from
the era when the pregalaxies were made of the cosmic strings. This
is possible due to the scaling\cite{scaling} property for the
cosmic string network, but for a more concrete conclusion we need
numerical studies.

On the other hand, if the cosmic sting were formed when the $U(1)$
symmetry was broken at extremely high energy near the
string-unification scale($\sim 5\times 10^{17}$ GeV),\cite{ms} we
could have the constant value of rotation velocity comparable to
the measured one\cite{flat} in the galaxy, which is including the
remnants of the single string. However, for the explanation to be
valid for flat rotation curves we should consider scalar-tensor
gravity theories with $a\neq -1$ as we can see in Eq. (3.28). A
simple model for the string-seeded galactic halo with two-dilaton
fields has been constructed in the last section, with the case
$\delta=0$ considered only. Two-dilaton fields without their
potential terms such as Eq. (4.1) have been considered elsewhere,
for example, in a two-dilaton model of electroweak
interactions.\cite{2D}  The case\cite{axion} $\delta\neq 0$ can be
similarly studied, but the solution of the dilaton field $\phi(r)$
for $r<r_h$ seems not to be expressed in elementary functions. Our
consideration in this paper has been restricted to the static
spacetime with cylindrical symmetry, and it seems interesting to
generalize ours to the cylindrical spacetime with
rotation\cite{Santa}.

\acknowledgments

The author would like to thank his family for their helps. This
work was supported by the Soongsil University Research Fund.

\end{document}